\documentclass[twocolumn,floatfix,english]{revtex4}
\usepackage{pifont}
\usepackage{dcolumn}
\usepackage{amsmath}
\usepackage{amssymb}
\usepackage{graphicx}
\usepackage{bm}
\usepackage[utf8]{inputenc}
\usepackage[T1]{fontenc}
\usepackage{color}
\usepackage{url}
\usepackage[bf]{subfigure}
\usepackage{rotating}
\usepackage{mathrsfs}
 \usepackage{multirow}
 \usepackage{longtable}
\usepackage{url}

\usepackage{lipsum, babel}

\begin{document}

\title{The effect of interurban movements on the spatial distribution of population in China}

\author{Jiachen Ye,$^{1,2}$ Qitong Hu,$^{1,2}$ Peng Ji,$^{1,2}$}
 \email{pengji@fudan.edu.cn}

\author{Marc Barthelemy$^{3,4}$}
 \email{marc.barthelemy@ipht.fr}

\affiliation{$^1$Institute of Science and Technology for Brain-Inspired Intelligence, Fudan University, Shanghai 200433, China}
\affiliation{$^2$Research Institute of Intelligent and Complex Systems, Fudan University, Shanghai 200433, China}
\affiliation{$^3$Institut de Physique Th\'eorique, Universit\'e Paris Saclay, CEA, CNRS, F-91191 Gif-sur-Yvette, France}
\affiliation{$^4$Centre d'Analyse et de Math\'ematique Sociales, (CNRS/EHESS) 54, Boulevard Raspail, 75006 Paris, France}

\date{}


\begin{abstract}
  Understanding how interurban movements can modify the spatial distribution of the population is important for transport planning but is also a fundamental ingredient for epidemic modeling. We focus here on vacation trips (for all transportation modes) during the Chinese Lunar New Year and compare the results for 2019 with the ones for  2020 where travel bans were applied for mitigating the spread of a novel coronavirus (COVID-19). We first show that these travel flows are broadly distributed and display both large temporal and spatial fluctuations, making their modeling very difficult. When flows are larger, they appear to be more dispersed over a larger number of origins and destinations, creating de facto hubs that can spread an epidemic at a large scale. These movements quickly induce (in about a week) a very strong population concentration in a small set of cities. We characterize quantitatively the return to the initial distribution by defining a pendular ratio which allows us to show that this dynamics is very slow and even stopped for the 2020 Lunar New Year due to travel restrictions. Travel restrictions obviously limit the spread of the diseases between different cities, but have thus the counter-effect of keeping high concentration in a small set of cities, a priori favoring intra-city spread, unless individual contacts are strongly limited. These results shed some light on how interurban movements modify the national distribution of populations, a crucial ingredient for devising effective control strategies at a national level.
\end{abstract}




\maketitle




\section{Introduction}

\indent In early January 2020, we observed the outbreak of a novel coronavirus (COVID-19) in Wuhan, China, which has been quickly spreading out to the whole country, and more recently to other countries in the world \citep{WHO}. Infectious diseases spread among humans because of their interactions and movements, and the proximity of this outbreak with the Spring Festival, a period of travel with high traffic loads, provided terrible conditions for the spread of this disease. With an increasing amount of confirmed cases, more attention has been devoted to modeling the spread of COVID-19 from various aspects such as determining the value of the reproductive number  \cite{cao2020estimating, tang2020estimation, riou2020pattern, zhao2020preliminary, park2020reconciling, Zhang2020evolving}, of the incubation period  \citep{backer2020incubation, read2020novel, liu2020transmission}. In general, analytical modeling plays of course an important role in the prediction of the spread and allows in particular to test control strategies \citep{Heesterbeek2015modeling}, which was verified in this case too \citep{liang2020simple, chen2020time, ming2020breaking, majumder2020early, chen2020sars, li2020robust,wu2020nowcasting, chinazzi2020preliminary, lai2020preliminary, chinazzi2020effect, lai2020preliminary, chinazzi2020effect, qian2020scaling}. Particularly important were estimate of probability to export the disease in other countries \citep{pullano2020novel, chinazzi2020preliminary, gilbert2020preparedness}, and how effective were travel restrictions inside China \citep{chinazzi2020preliminary}.

Demographic information and mobility, either under the form of data or given by transportation models (see for example the review \citep{barbosa2018human}),  are crucial for these transmission models. Mobility can concern either the global scale with movements between countries  \citep{wu2020nowcasting, chinazzi2020preliminary, lai2020preliminary, chinazzi2020effect}, or the national scale between cities, or even inside cities \citep{lai2020preliminary, chinazzi2020effect, qian2020scaling}. In this study we are focusing on the national level and we won't try to model the spread of the disease. Instead we will focus on the statistical properties of the interurban mobility, and how it affects the spatial distribution of populations. More precisely, we will investigate the statistical properties, of traffic flows between cities during the Chinese Spring Festival in 2020 and by comparing with data for 2019, which are the most salient differences induced by travel restrictions. This knowledge will help us to understand the effect of travel restrictions and their impact on epidemic spread, and more generally to guide us for modeling mobility at this scale, a crucial ingredient in epidemiological studies, but also for other fields such as transportation planning.

\section{Statistics of interurban flows}

We will first study standard statistical properties of interurban flows obtained from migration data collected by
Baidu Qianxi (see Material and Methods). This dataset enables us to monitor the traffic flows between cities. For each day \(d\) (\(d=1,2,\dots,T\)), we can extract the number of individuals $N(i,j,d)$ going from city $i$ to city $j$ with any
travel mode. The migration data can thus be seen as a directed, weighted network of flows between the set of $n=296$ cities of China whose populations are also known (see Material and Methods).  We collected the data for the Spring Festival of 2020 (from Jan. 1st to Feb. 12th, 2020) and for assessing the impact of travel bans, we also collected the data for the Spring Festival of 2019 (which according to the Chinese lunar calendar takes place from Jan. 12th to Feb. 23rd, 2019).

\subsection*{Large heterogeneity of flows}

 \begin{figure}
   \centering
   \includegraphics[width=0.4\textwidth]{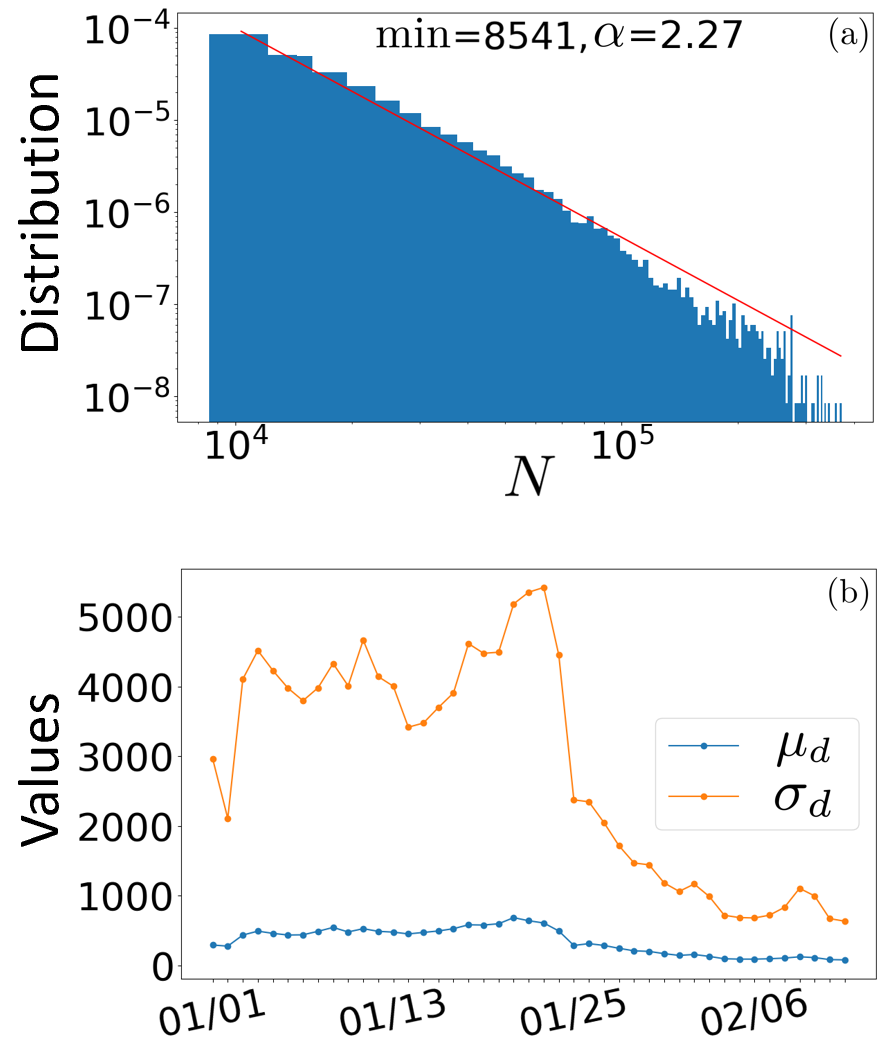}
   \caption{\small (a) Distribution of all traffic flows $N(i,j,d)$ in loglog. The line is a power law fit of the form $N^{-\alpha}$ with exponent $\alpha=2.27$ with fitting method described in \cite{Alstott2014powerlaw}. (b) Average and standard deviation of the flows $N(i,j,d)$ averaged over traffic flows versus the date $d$ (from 1st January to 12th February).}
   \label{Fig_distribution_traffic_flow}
 \end{figure}

We first consider the distribution of all flows of individuals $N(i,j,d)$ for all cities $i$ and $j$ and all days $d$ and which is shown in Fig.~\ref{Fig_distribution_traffic_flow} (a). The maximum flow is of order $10^5$ and the average of order $10^3$ indicating a broad distribution. A power law fit is consistent with this picture with an exponent $\alpha\approx 2.3$ (Fig.~\ref{Fig_distribution_traffic_flow} (a)). This heterogeneity is confirmed in Fig.~\ref{Fig_distribution_traffic_flow} (b) which shows both the average value \(\mu_{d}\) and the standard deviation \(\sigma_{d}\) computed over all inter-city flows (for each day $d$). We see that for most days the relative dispersion $\sigma_{d}/\mu_d$ is of order $5-10$. This heterogeneity is probably due to the large diversity of cities that can serve as origins or destinations of flows (see below for further analysis). An important feature that we can observe on Fig.~ \ref{Fig_distribution_traffic_flow} (b) is the sharp drop of the standard deviation after Jan. 25th, the Lunar New Year (LNY), which we will see below is mainly due to the travel ban (see also Fig.~S1, S3 in SI for a detailed discussion).

 \subsection*{Temporal versus spatial fluctuations}

\begin{figure}
   \centering
   \includegraphics[width=0.5\textwidth]{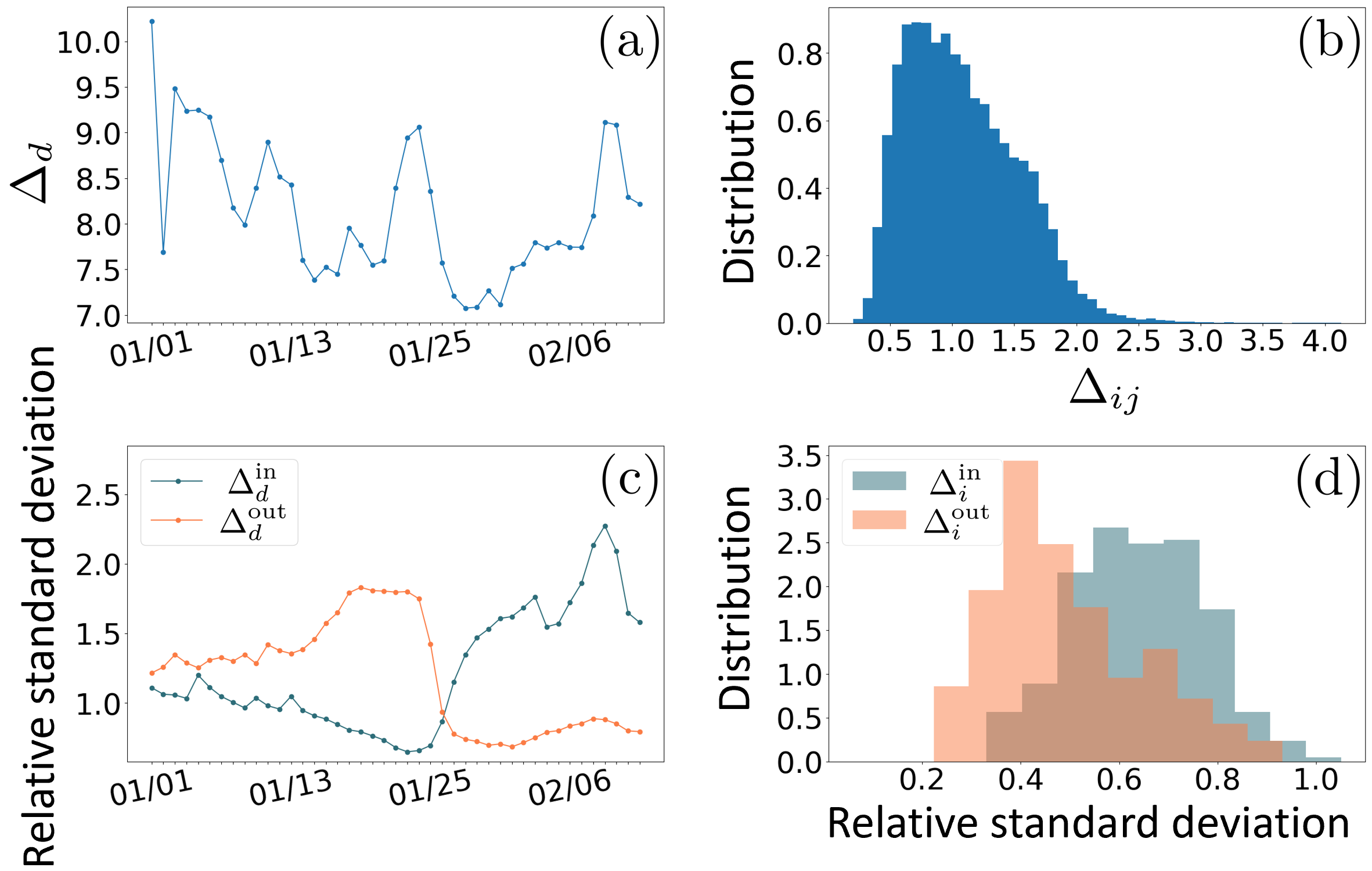}
   \caption{\small  (a) Relative standard deviation of the flows \(N\) averaged over traffic flows and represented here versus time. (b) Distribution of the relative standard deviation of \(N\) averaged over time. (c) Relative standard deviation of  $N_\text{in}(i,d)$ and $N_\text{out}(i,d)$ averaged over cities and shown here versus time. (d) Distribution of the relative standard deviation of \(N_\text{in}\) and \(N_\text{out}\) averaged over time}
   \label{Fig_temporal_spatial}
 \end{figure}

 In order to understand the nature of the different fluctuations affecting the flows $N(i,j,d)$, we compute the relative standard deviation \(\Delta_{ij}= \frac{\sigma_{ij}}{\mu_{ij}}\), where \(\mu_{ij}\) and \(\sigma_{ij}\) are the average and standard deviation computed over time, and the relative standard deviation \(\Delta_{d}=\frac{\sigma_{d}}{\mu_{d}}\), average over all flows, for a given day $d$. We show on Fig.~\ref{Fig_temporal_spatial} (a) the spatial dispersion  $\Delta_{d}$ versus time and in Fig.~\ref{Fig_temporal_spatial} (b) the distribution of  $\Delta_{ij}$. We observe that the spatial dispersion is of order $8.3$, while the temporal dispersion is less (mainly concentrated around $1$). The main reason for heterogeneity thus lies in the flow fluctuations between different origins and destinations, while temporal fluctuations are smaller but not negligible. These two sources of heterogeneity clearly represent a challenge for modeling these flows, especially with very simplified models.
Our results indicate that the first modeling step would be to describe the spatial heterogeneity of flows and then consider temporal variations.

 The next natural quantities that can be computed over this network are the incoming and outgoing flows defined by
 \begin{align}
 \begin{cases}
 N_\text{in}(i,d) = \sum_{j=1}^nN(j,i,d)\\
 N_\text{out}(i,d) = \sum_{j=1}^nN(i,j,d)
\end{cases}
\end{align}
respectively. We measure in the same way as above various measures of fluctuations, either averaged over cities or over time, leading to the quantities $\Delta_{d}^{\text{in (out)}}$, $\Delta_{i}^{\text{in (out)}}$. As these quantities are sums of random variables, we expect smaller relative dispersions than for $N(i,j,d)$ which is indeed what we observe (see Fig.~\ref{Fig_temporal_spatial} (c) and (d), with typical values of relative dispersion of order $1$ (see Figs. S4,S5 for additional details). In order to get first insights about the influence of travel bans, we compare the incoming flows and outgoing flows versus city population in 2019 and 2020 with days $N_{\text{in, out}}^{\text{before}}$ before and $N_{\text{out,in}}^{\text{after}}$ after LNY. We first observe that
(see Fig.~S2 in Supplementary Information (SI))  basically the number of outgoing individuals before LNY corresponds approximately to the number of incoming individuals after LNY with  $N_{\text{in (out)}}^{\text{before}}\approx N_{\text{out (in)}}^{\text{after}}$  (and vice-versa). These relations thus correspond roughly to the conservation of the number of individuals traveling during the Chinese Spring Festival.

 \subsection*{Structure of incoming and outgoing flows}

 The value of incoming or outgoing flows gives information about the volume of migrations, but not about the number of important origins or destinations. In order to characterize the dispersion over different cities, we denote by \(\mathcal{O}(i,d)\) and \(\mathcal{D}(i,d)\), the sets of origin of flows incoming in city $i$ and destinations of flows from city $i$ (for the day $d$), respectively. We then use Gini indices \citep{dixon1987} that capture the dispersion of incoming and outgoing flows and are given by
\begin{align}
  G_\text{in}(i,d) = \frac{1}{2O^2\overline{N}_\text{in}(i,d)}\sum_{p,q \in \mathcal{O}(i,d)} |N(p,i,d)-N(q,i,d)|\\
  G_\text{out}(i,d) = \frac{1}{2D^2\overline{N}_\text{out}(i,d)}\sum_{p,q \in \mathcal{D}(i,d)} |N(i,p,d)-N(i,q,d)|
\end{align}
where $O$ and $D$ represent the number of elements of the sets  \(\mathcal{O}(i,d)\) and \(\mathcal{D}(i,d)\). The quantity  \(\overline{N}_\text{in}(i,d)=\frac{N_\text{in}(i,d)}{|\mathcal{O}(i,d)|}\) is the average incoming flows and \(\overline{N}_\text{out}(i,d)=\frac{N_\text{out}(i,d)}{|\mathcal{D}(i,d)|}\) the average outgoing flows.  Intuitively, if all traffic flows to city \(i\) are from one single origin city on day \(d\), the Gini index \(G_\text{in}(i,d)\) will be \(1\), while if traffic flows to city \(i\) are all equal, the Gini index \(G_\text{in}(i,d)\) will be \(0\) (and similarly for \(G_\text{out}(i,d)\)).

\begin{figure}
   \centering
   \includegraphics[width=0.5\textwidth]{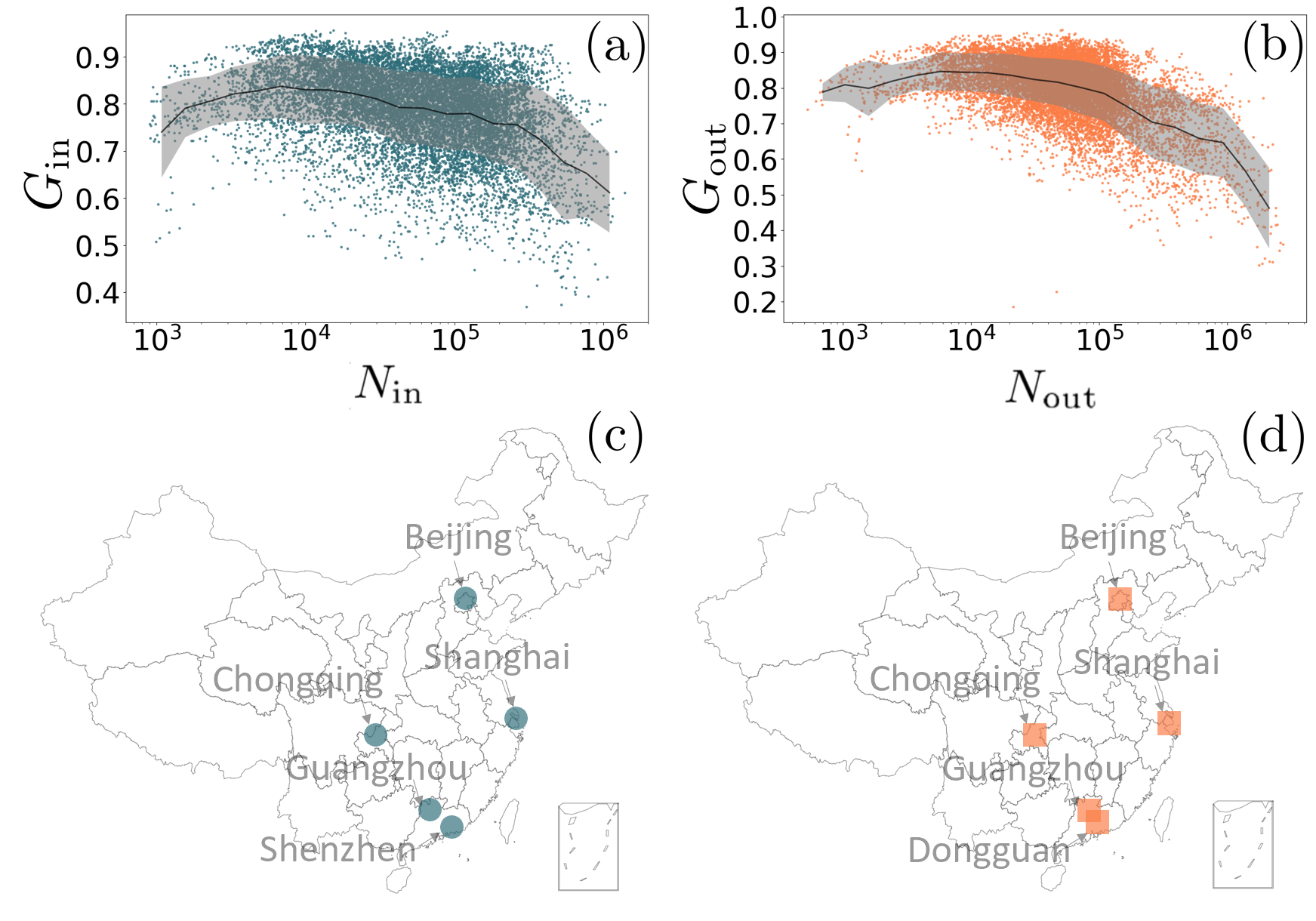}
   \caption{\small (a) \(G_\text{in}\) versus \(N_\text{in}\) for all cities and days.
   (b) \(G_\text{out}\) versus \(N_\text{out}\) for all cities and days (shown in loglog).
   The thick line indicates the average of  \(G_\text{in}\)  ( \(G_\text{out}\) ) versus of $N_{\text{in}}$ ($N_{\text{out}}$) and the shaded area represents the standard deviation of the average. Top 5 critical cities for (c) incoming flows and (d) outgoing flows.}\label{Fig_gini_index_cities}
 \end{figure}

 We plot these Gini indices computed for each city versus the traffic flows to or from this city. These figures \ref{Fig_gini_index_cities} (a,b) show that on average the larger the traffic flows are, the more dispersed they are over a larger number of origins or destinations. In terms of epidemic control, it is clear that cities with a large flow \(N_\text{in}\) and a small Gini index \(G_\text{in}\) are the most critical, in the sense that many people from many different cities are converging to the same place. Equally, cities with a large \(N_\text{out}\) and a small \(G_\text{out}\) should be particularly monitored, since they can act as hubs in spreading the disease over the inter-city network. As shown in Fig. \ref{Fig_gini_index_cities} (c,d), we show the top $5$ critical cities,
including  Beijing, Shanghai, Chongqing and Guangzhou for  both the incoming and outgoing flows, Shenzhen for the  incoming flows, and Dongguan for the outgoing flows.

\section{Statistical structure of the national population}

An important effect of incoming and outgoing flows is that they change the population structure. Some cities will receive a large number of individuals while for others we expect a decrease of their population. Migration thus affects the statistical structure of the national population and in this section we will characterize this effect.

\subsection*{Temporal evolution of population structure}

\begin{figure}
   \centering
   \includegraphics[width=0.5\textwidth]{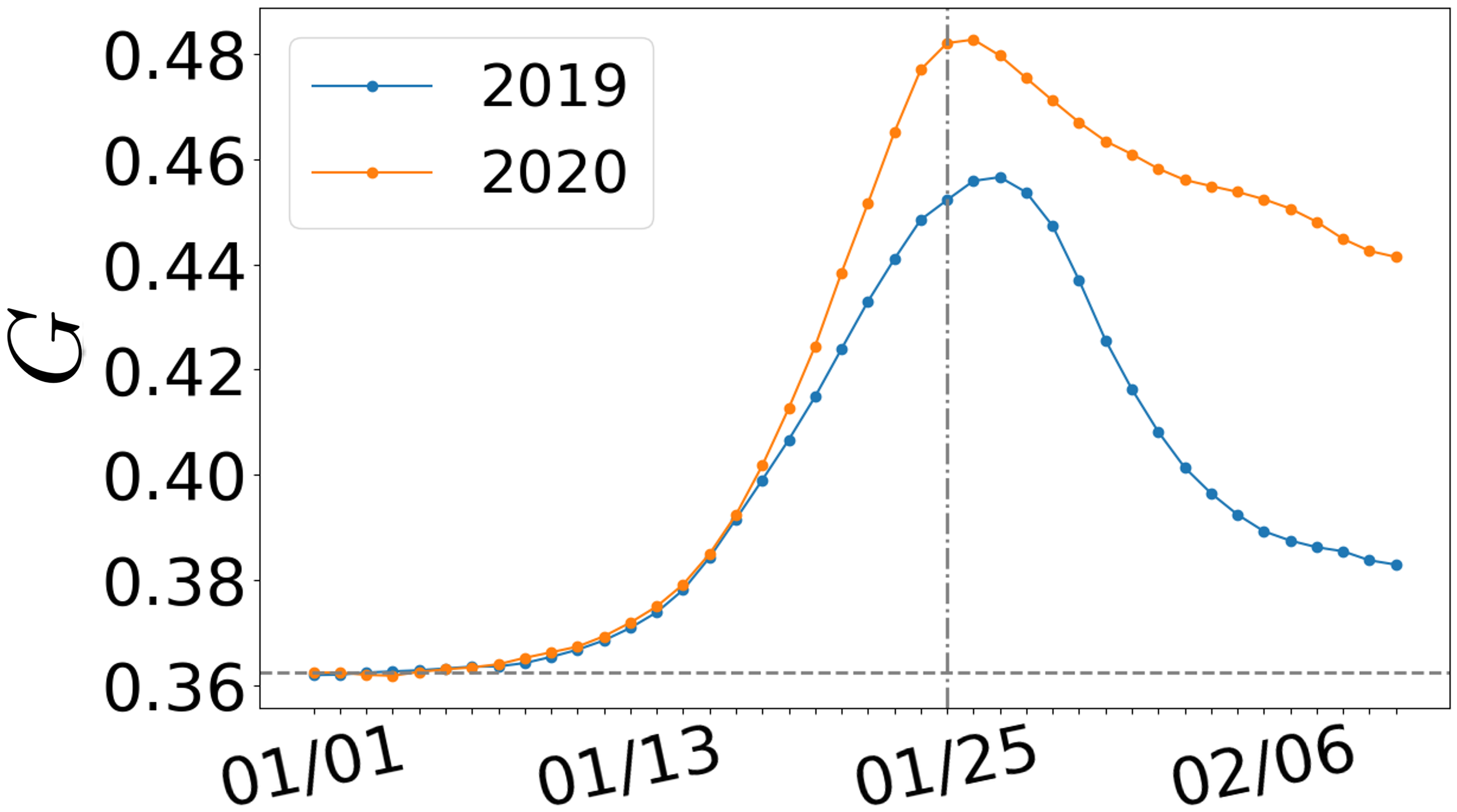}
   \caption{\small Temporal variations around the Spring Festival holidays of the population Gini index for 2019 and 2020. The horizontal dotted line represents the value `at rest'. The vertical line indicates the day of the LNY.} \label{Fig_gini_t}
 \end{figure}

In order to characterize the disparity of the population distribution and how it varies during seasonal migrations, we consider the population of city $i$ at time $d$ given by
\begin{align}
P(i,d) = P_0(i)+\sum_{d'\leqslant d}N_\text{in}(i,d')-\sum_{d'\leqslant d}N_\text{out}(i,d'),
\end{align}
where \(P_0(i)\) represents the population of city \(i\) without incoming and outgoing flows. The Gini index for the city population of the whole country at day \(d\) is then given by
\begin{align}
   G(d) = \frac{1}{2n^2\overline{P}(d)}\sum_{i,j=1}^n |P(i,d)-P(j,d)|,
 \end{align}
where \(\overline{P}(d) = \frac{1}{n}\sum_{i=1}^nP(i,d)\) is the average population of all cities at day \(d\). Intuitively, if all people gather in one city, \(G\) will be \(1\), while if people spread evenly across all cities, \(G\) will be \(0\). For comparison, we also define the Gini index at rest as
 \begin{equation}
 G_\text{rest} = \frac{1}{2n^2\overline{P}_0}\sum_{i,j=1}^n |P_0(i)-P_0(j)|.
 \end{equation}
This quantity captures the degree of population concentration without any traffic flows, where \(\overline{P}_0=\frac{1}{n}\sum_{i=1}^nP_0(i)\) is the average population of all cities without any traffic flows. We show in Fig.~\ref{Fig_gini_t} the variation of the Gini coefficient when we take into account migration flows.
 We plot both the results for 2019 and 2020. In both cases we see an important increase of the Gini coefficient in a short time (about a week): when the LNY is approaching, people go back from workplaces to hometowns for reunion with families. A smaller set of cities concentrates these meetings with the number of important cities reaching its minimum and the Gini index reaching its peak on the LNY. Based on the Gini index, we can estimate the number of `important' cities where the concentration takes place through \([n(1-G)]\), where \([\cdot]\) denotes the integer part (see SI for details where we show in Fig.~S6 this number versus time and we indeed observe an important drop when approaching the LNY). After the LNY (Jan. 25th), individuals are going back home and the Gini coefficient relaxes back to its original value, but much slower. We observe that in 2020, the increase of the Gini index is larger and, due to travel bans, the decrease even slower than normal. The reason may be that after the outbreak of COVID-19, almost all regions have deferred the time of resuming works and classes after the Spring Festival holiday. For example, Shanghai proposed that companies not crucial to the nation should not resume works before Feb. 10th and that schools should provide online classes. At this point, the population structure at the national level is far from being back to normal. These different results show that these seasonal movements induce a strong concentration of individuals in a relative small set of cities, and that travel bans tend to keep this situation of high concentration.

\begin{figure}
   \centering
   \includegraphics[width=0.5\textwidth]{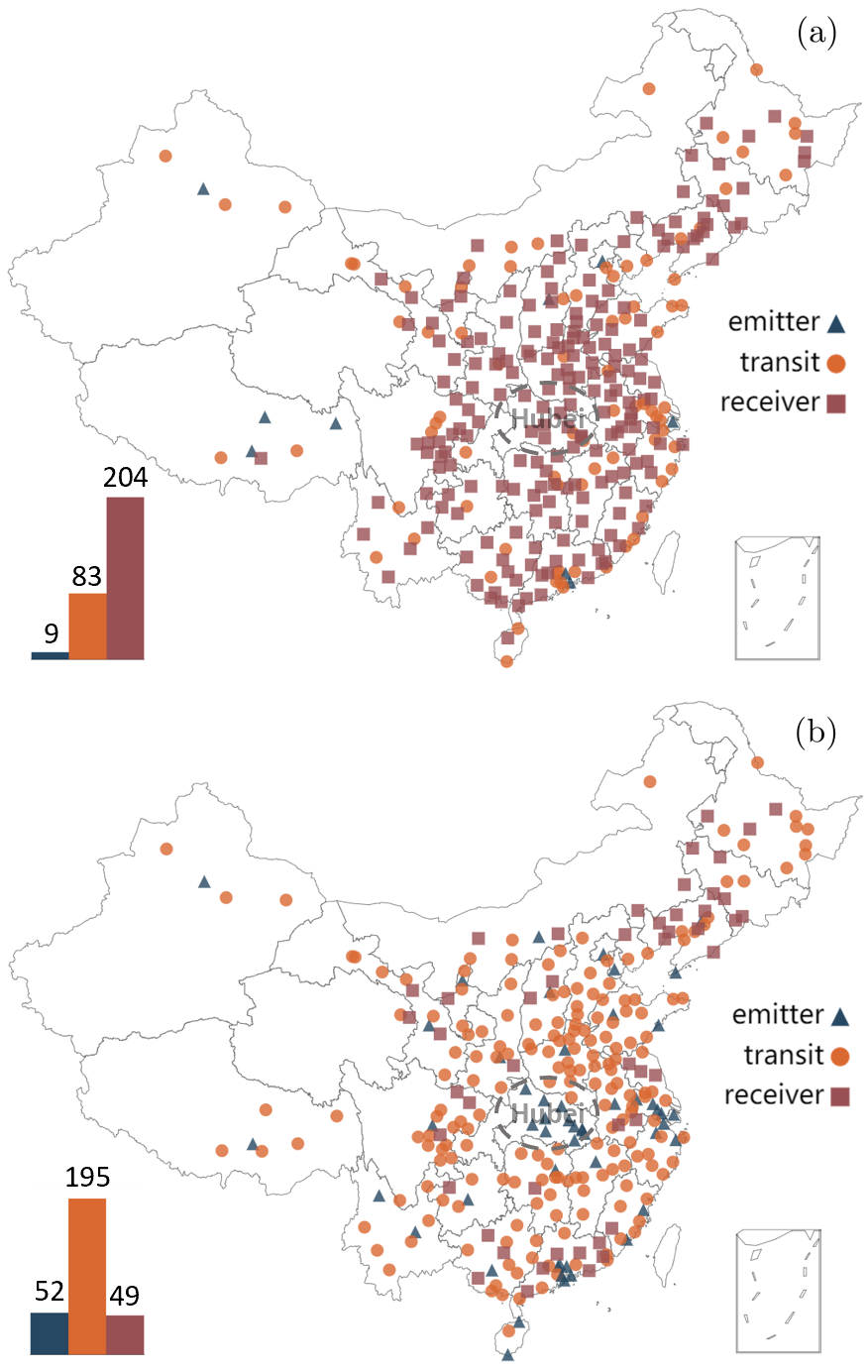}
   \caption{\small (a) Emitter, receiver and transit cities according to the value of \(R(i,1)\) for 2019 with the number of three categories of cites at lower left corner. (b) Emitter, receiver and transit cities according to the value of \(R(i,1)\) for 2020 with the number of three categories of cites at lower left corner.}
   \label{Fig_average pendular ratio on map}
 \end{figure}

\subsection*{Return to `equilibrium': pendular ratio}

We observe in Fig.~\ref{Fig_gini_t} that after the LNY there is a decrease of the Gini index indicating a return to normal state characterized by a lower concentration of individuals. In order to characterize quantitatively this return to the original state (before holidays), we measure the gap between individuals going out from a city before the LNY and coming back after it. This gap defines a `pendular ratio' given by
 \begin{align}
 R(i,d_f) = \frac{\sum_{\hat{d}< d \leqslant \hat{d}+d_f}N_\text{in}(i,d)}{\sum_{\hat{d}-d_f \leqslant d<\hat{d}}N_\text{out}(i,d)},
\end{align}
where \(d_f\) is a range of days around the LNY \(\hat{d}\). If this ratio is much larger than 1, it means that for this city there is a large incoming flow while for the opposite situation $R(i,d_f)\ll 1$, a large number of individuals are going out (compared to the incoming flows). At large times $d_f$, we expect that $R\simeq 1$ since most of the individuals have come back. We divide cities into three categories according to the value of \(R(i,1)\): If the value is larger than \(1.5\), we classify city \(i\) as a `receiver' city. If the value is less than \(0.5\), we classify city \(i\) as an `emitter' city. Finally, if the value is between \(0.5\) and \(1.5\), we classify city \(i\) as a `transit' city. We represent on Fig.~\ref{Fig_average pendular ratio on map} the cities of different types on the map of China. We observe that both receiver and transit cities are homogeneously distributed in China. In constrat emitters cities are in general located in developed regions, e.g., Beijing, Shanghai, Guangzhou, and so on, as shown in figures \ref{Fig_average pendular ratio on map} (a) and (b). It is interesting to note that cities of the Hubei province (within the dashed circle in the figure) are emitters cities in 2020, essentially due to travel restrictions that prevented individuals to come back to Wuhan. This is an important difference compared to the year of 2019 that appears here in the spatial structure of emitters and receivers. 

\begin{figure}
   \centering
   \includegraphics[width=0.5\textwidth]{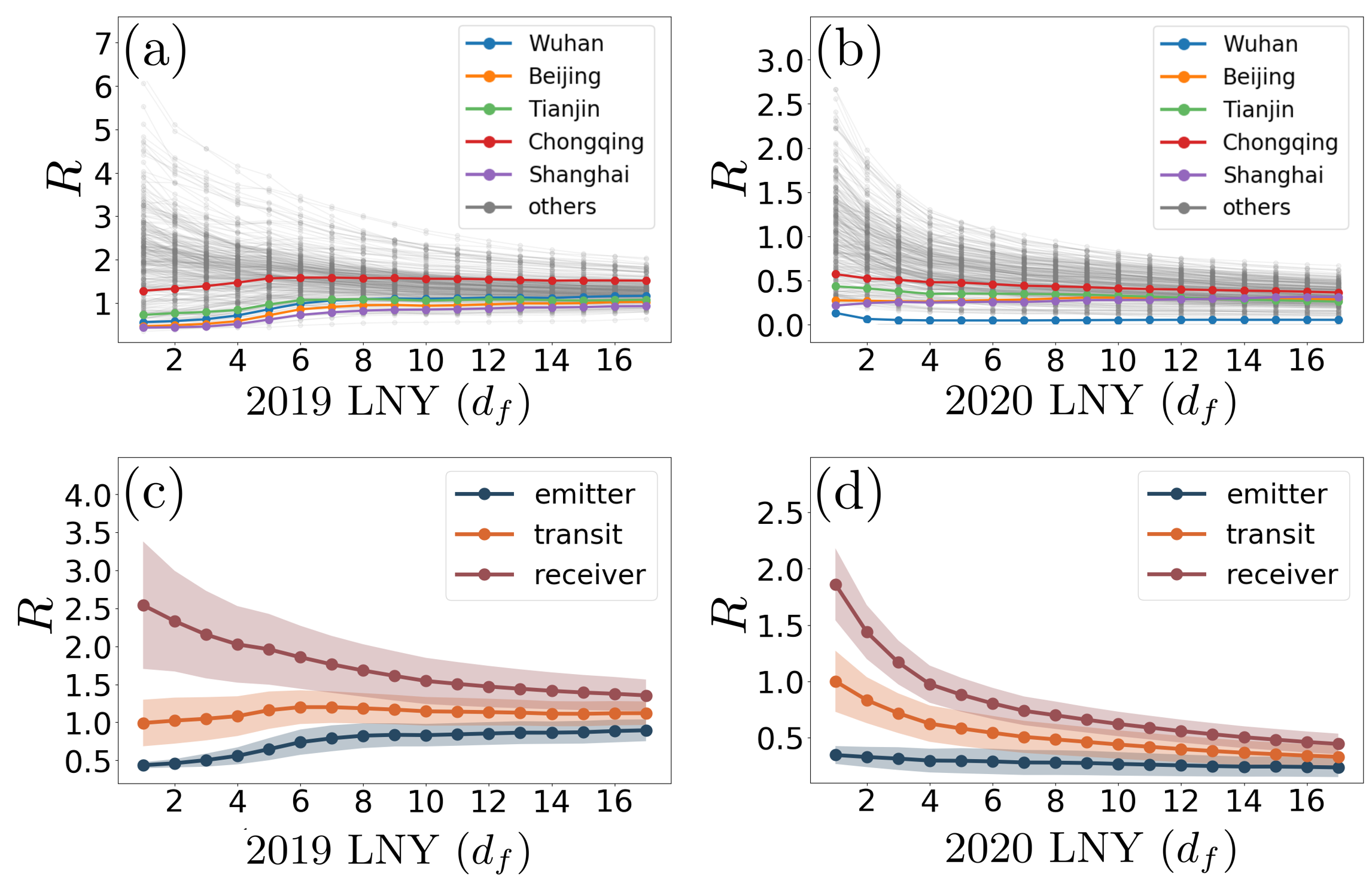}
   \caption{\small Comparison of pendular ratio between 2019 and 2020. (a) Pendular ratio for all cities versus days $d_f$ from LNY in 2019. We highlight five important cities. (b) Pendular ratio for all cities versus days $d_f$ from LNY in 2020 with highlight of 5 cities. (c) Average values of the pendular ratio over cities according to the classification (receiver, emitter or transit cities) versus days from LNY in 2019. The colored areas correspond to one standard deviation. (d) Mean value of the pendular ratio over cities according to the classification versus days from LNY in 2020 with the shaded area representing the corresponding standard deviation.}\label{Fig_pendular}
 \end{figure}

 We show in Fig.~\ref{Fig_pendular} (a,b) the pendular ratio for 2019 and 2020 for all cities and we highlight 5 cities: Wuhan, Beijing, Tianjin, Chongqing and Shanghai, corresponding to the origin place of COVID-19 and four province-level municipalities. We note here that the curve corresponding to Wuhan is at the bottom of all cities in Fig.~\ref{Fig_pendular} (b), reflecting the success of sealing off Wuhan from all outside contact to stop the spread of the disease since Jan. 23rd.

 In Fig.~\ref{Fig_pendular} (c,d) we show this pendular ratio for 2019 and 2020 for  the different types of cities (we average over cities in a given category, emitter, receiver or transit). We observe that the standard deviation is small for the three groups adding credit to their definition. In addition, compared to 2019, the values of \(R(i,1)\) corresponding to 2020 are much  smaller. We observe that in 2019, the pendular ratio of all the three types of cities returns to 1, meaning that the majority of individuals who went away for the holidays came back. The situation for 2020 is very different with a pendular ratio for all types of cities that converges to a value less than 1 (even less than $0.5$), indicating that the majority of people who went away for the holidays did not come back yet. This result remains consistent with the conclusion of Gini index (Fig.~\ref{Fig_gini_index_cities}) about a larger concentration in cities and the effect of travel bans.

 Finally, we note here that we additionally implemented our whole analysis at the province level (see SI) and the results obtained are similar are those obtained at the city level.

\section{Conclusion}

Our findings thus concern four different aspects. First, the traffic flows between cities are very heterogeneous not only spatially but also from a temporal perspective. Such a large heterogeneity could be induced by both the Spring Festival and the travel ban. Similar results apply also to an aggregated level, i.e. the incoming and outgoing flows for cities also display important heterogeneities. This aspect is crucial for understanding and modeling epidemic spreading for which we know the importance of heterogeneity for epidemic spreading on networks \cite{pastor2001,barthelemy2004} and more generally for most processes \cite{barrat2008}. We also quantify the dispersion of origins/destinations of the incoming/outgoing flows showing that for larger flows we have a larger variety of origins and destinations. We also show that during these seasonal migrations of the Spring Festival, the national structure of population changes quickly with a larger concentration in a small set of cities. This concentration decays normally in time after the festivities but travel bans slow down this return to the initial state. It is natural to try to stop the geographical spread of the disease by stopping interurban movements, but on the other hand, large concentration in cities can favor the spread at the city level and increase the number of infected cases. This concentration can be compensated by a more important control at the individual contact level which is what was done in cities such as Wuhan. These results are in line with epidemic modeling results \citep{chinazzi2020preliminary}, where it was shown that travel quarantine is effective only when combined with a large reduction of intra-community transmission. Our results thus highlight the importance of mobility studies for modeling a variety of processes and in particular for understanding and modeling the spread of epidemics. Effective mitigating strategies need to take into account the change of population structure that we exhibited here.

\section{Material and methods}
\subsection*{Data}
 \indent We obtained the migration data from Baidu Qianxi (http://qianxi.baidu.com), based on Baidu Location Based Services and Baidu Tianyan, for all transportation modes. It provides the following two datasets: migration index reflecting the size of the population moving into or out from a city/province, and migration ratio capturing the proportion of each origins and destination. We collected the data during Chinese Spring Festival period of 2020 (from Jan. 1st to Feb. 12th, 2020). For parallel comparison, the migration index during the same period of 2019 (re-scaled according to Chinese lunar calendar, from Jan. 12th to Feb. 23rd, 2019) is also used.

 In addition to the migration data, we collected the demographic from China Statistical Yearbook (http://www.statsdatabank.com), an annual statistical publication, which reflects comprehensively economic and social development of China. It covers key statistical data in recent years at both the city level and the province level. We collected the data of population of 31 province-level regions and 296 city-level regions from China Statistical Yearbook 2019, the latest edition provided.

\bibliographystyle{unsrt}
\bibliography{2019-nCov}

\pagebreak 
\newpage

\onecolumngrid
\appendix

\section{Supplementary Materials}

\setcounter{equation}{0}
\setcounter{figure}{0}
\renewcommand{\thefigure}{S\arabic{figure}}
\renewcommand{\theequation}{S\arabic{equation}}

\subsection*{Data for 2019}

In order to evaluate the heterogeneity of flows of 2019 with comparison to that of 2020, we use the  migration index during the same period of 2019 (re-scaled according to Chinese lunar calendar, from Jan. 12th to Feb. 23rd, 2019),  and we would compute the distribution of all flows $N(i,j,d)$, for all cities $i$ and $j$ and all days $d$, though \( N(i,j,d) = N_\text{\text{out}}(i,d) \times p(i,j,d)\). The Chinese Lunar New Year of 2019 is Feb. 5th. Here, $N_\text{\text{out}}(i,d) $ is  migration index reflecting the size of the population moving into or out from a city/province, and $p(i,j,d)$ is  migration ratio capturing the proportion of each origins and destination. However, the migration ratio is unavailable for 2019.
We apply the data of \(p(i,j,d)\) for 2020 to the computation of \(N(i,j,d)\)  for 2019, with results shown in Fig.~S1. This result exhibits large heterogeneity of flows  and  displays a localized drop around LNY.

\subsection*{Statistics of $N_{\text{\text{in}}}$ and $N_{\text{\text{out}}}$}

\subsubsection*{Versus population}

We observe a power law relationship between the incoming/outgoing flows and city population in Fig.~S2 which indicates that the larger a city and the more flows it carries. Compared to 2019, the differences between the scatter points for incoming flows corresponding to days before and after LNY are much larger in 2020. This result emphasizes again that travel ban causes indeed the sharp drop of standard deviation in Fig.~1 (b) of the main text rather than the low travel intention during the Spring Festival.

The power law fits that we obtain imply that $N_{\text{\text{in}}}\sim P_0^{\gamma_{\text{in}}}$ where $\gamma_{\text{in}}\approx 0.93$ before LNY and $\gamma_{\text{in}}\approx 0.88$ after LNY. Similarly for outgoing flows we obtain $\gamma_{\text{out}}\approx 0.85$ before and $\gamma_{\text{out}}\approx 0.93$ after LNY. We can interpret these results as a consequence of the conservation of the number of individuals traveling before and after LNY.

\subsubsection*{Distribution of $N_{\text{in,out}}$ for 2020}

We show the distribution, average and standard deviation of incoming/outgoing flows in Fig.~S3. We observe that these distributions are relatively broad, in particular outgoing flows (Fig.~S3 (a) and (b)).
We show the standard deviation of incoming flows and outgoing flows over cities for each day, and the corresponding average (the same for the incoming  and outgoing flows) in Fig.~S3 (c). Note that the standard deviation of \(N_\text{\text{in}}\) is smaller than that of \(N_\text{\text{out}}\) before Jan. 25th, the 2020 LNY, while the situation reverses after Jan. 25th. A reasonable explanation for this is that people go to a relative large number of hometowns from a relative small number of workplaces before the Spring Festival; Due to the travel ban, people do not come back after the Spring Festival.

\subsubsection*{Distribution of $N_{\text{in,out}}$ for 2019}

We show the same quantities as above but for the year 2019. Here also, we observe broad distributions both for  all incoming flows and all outgoing flows are shown in Fig.~S4.

Fluctuations are larger around LNY where the total flow of  individuals is larger, allowing for more heterogeneity. Before LNY individuals move from a large variety of  cities to a relatively small number of hometowns explaining the large fluctuations of $N_{\text{out}}$. After LNY, individuals are returning from a small number of hometowns to a large variety of cities, inducing large fluctuations of $N_{\text{in}}$. The corresponding dispersion and relative dispersions $\Delta^{\text{out}}_{d}$) and $\Delta^{\text{in}}_{d}$) are shown in figures~S4 (c) and (d). This also results in the heterogeneous distribution for the relative standard deviation of \(N_\text{\text{in}}\) and \(N_\text{\text{out}}\) averaged over cities for 2019 in Fig.~S4 (e).

\subsubsection*{Gini indices for $N_{\text{in,out}}$}

We compute Gini indices for cities. Instead of showing results for all cities, we plot Gini indices versus the traffic flows to or from cities in set \(\{i \in \mathcal{V} | \mathop{\text{min}}\limits_{d}\{\mathop{\text{max}}\limits_{d}\{\frac{N_\text{\text{in}}(i,d)}{N_\text{\text{out}}(i,d)}\},\mathop{\text{max}}\limits_{d}\{\frac{N_\text{\text{out}}(i,d)}{N_\text{\text{in}}(i,d)}\}\}>4.5 \} \) and Wuhan in Fig.~S5. In this case from these cities, many people go out to or come in from many different cities. These cities are critical and include Shanghai, Beijing, and so on. Due to travel bans, Wuhan exhibits specific features of scatter points with clear separation before and after LNY (see the scatter points in blue at the upper left corner corresponding to Wuhan in Fig.~S5).

We also observe here in both cases a decreasing behavior on average. This is more salient for $N_{\text{out}}$ where the trend is clearly visible. This indicates that for larger outgoing flows, the Gini is smaller with no clearly dominant flow.

\subsection*{Statistical structure of the national population}

We first show the population distribution in Fig.~S6 (a). We observe a broad distribution and a power law fit gives the exponent \(\alpha \approx 5\). We also show the number of important cities quantified by the integer part of \(n[1-G(d)]\), in Fig.~S6 (b).

\subsection*{Pendular ratio}

In order to test the dependence of the pendular ratios on the criteria for defining classes, we change the criteria as follows: here if the value of \(R(i,1)\) is larger than \(1.2\), we classify city \(i\) as a `receiver' city. If the value is less than \(0.8\), we classify city \(i\) as an `emitter' city. Finally, if the value is between \(0.8\) and \(1.2\), we classify city \(i\) as a `transit' city. We show the location of three categories of cites on the map of China for 2019 and 2020 in Fig.~S7.

Compared to the criteria in the main text, the number of transit cities decreases, while the number of emitter and receiver cities increases. However, as shown in Fig.~S8, the patterns of the average value of pendular ratio corresponding to three categories of cites remain unchanged.

\subsection*{Statistics at the inter-province level}

\subsubsection*{Statistics of flows}

In what follows, we show some corresponding results based on province-level data instead of city-level data.
The distribution of traffic flows between provinces exhibits large heterogeneity with the exponent of power law fit \(\alpha\) around 2.24, as shown in Fig.~S9 (a). A sharp drop of the standard deviation after Jan. 25th is observed in Fig.~S9 (b).

We show the relative standard deviation of \(N\) over flows versus time with an order around 2.96 in Fig.~S10 (a) and the distribution of the relative standard deviation of \(N\) over time concentrating around 0.64 in Fig.~S10 (b). Large heterogeneity of traffic flows between provinces confirms the difficulty of modeling these flows. The relative standard deviations corresponding to incoming and outgoing flows with smaller relative dispersions are shown in figures~S10 (c) and (d).

We compare the incoming flows and outgoing flows versus city population in 2019 and 2020 at province-level with days before and after LNY highlighted by different colors in Fig.~S11. Compared to 2019 (figures~S11 (a) and (b)), the differences between days before and after LNY are much larger in 2020 (figures~S11 (c) and (d)).

\subsubsection*{Statistical structure of the national population}

The trends of the population Gini index for 2019 and 2020 are shown in Fig.~S12. The Gini index reaches its maximum around  LNY and returns to normal state gradually. Compared to 2019, the Gini index corresponding to 2020 has a higher peak and decreases with a slower speed.

We apply the criteria of three categories of provinces similar to that for city: If the value of \(R(i,1)\) is larger than \(1.2\), we classify province \(i\) as a `receiver' province. If the value is less than \(0.8\), we classify province \(i\) as an `emitter' province. Finally, if the value is between \(0.8\) and \(1.2\), we classify province \(i\) as a `transit' province. We show the location of three categories of cites on map of China for 2019 and 2020 in Fig.~S13 and observe that receiver provinces are the majority in 2019 while emitter provinces are the majority in 2020. This results seem to make sense since most people defer the return time due to the travel ban, so that most provinces are `emitters' in 2020.

We show in figures~S14 (a) and (b) the pendular ratio for 2019 and 2020 for all provinces and highlight 5 provinces: Hubei, Beijing, Tianjin, Chongqing and Shanghai, corresponding to the origin province of COVID-19 and four province-level municipalities. We note that the curve corresponding to Hubei is in the bottom from all provinces in Fig.~S14 (b), indicating that except Wuhan, the origin city of COVID-19, people also avoid going to  cities of Hubei. We observe that, in 2019, the pendular ratios of all the three types of cities return to 1, meaning that the majority of individuals who went away for the holidays came back, as shown in Fig.~S14 (c). The situation for 2020 is very different with a pendular ratio for all types of cities that converges to a value less than 1, indicating that the majority of people who went away for the holidays did not back yet, as shown in Fig.~S14 (d).

To sum up, we show that the traffic flows between provinces are very heterogeneous, and display both large temporal and spatial fluctuations, and so on. These results for province-level are in good agreement with  for city-level, indicating that our methods are applicable to both scales. Despite the detailed characters of traffic flows revealed by results for city-level and province-level, a global view of a higher level is also necessary.
The statistical properties of the interurban mobility help us to understand the effect of travel restrictions, their impact on and the control of epidemic spread.

 \begin{figure}[ht!]
   \centering
   \includegraphics[width=0.6\textwidth]{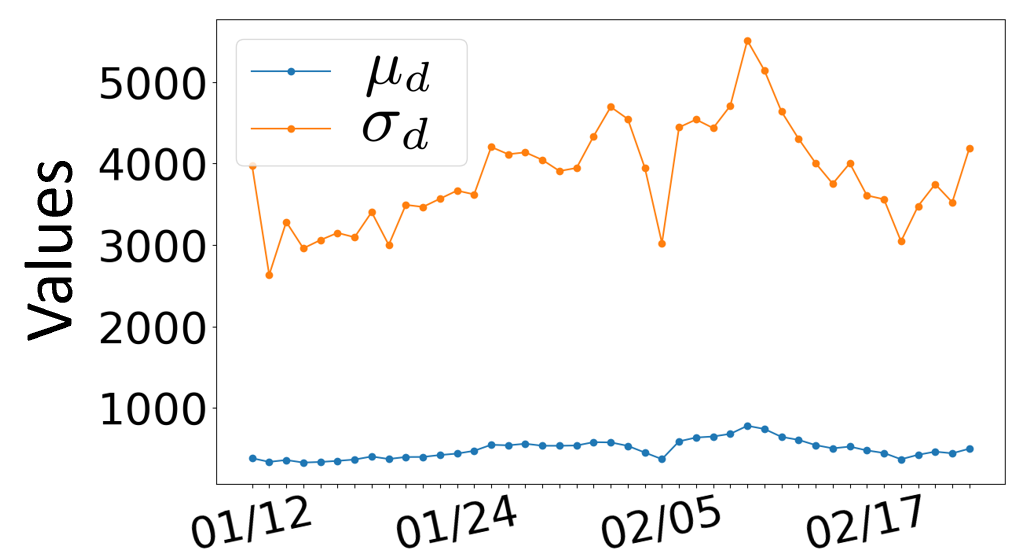}
   \caption{\small  Average and standard deviation of \(N\) over traffic flows versus time for 2019 with the corresponding migration ratio for 2020.}\label{Fig_average_standard_deviation_2019}
 \end{figure}

 \begin{figure}[ht!]
   \centering
   \includegraphics[width=0.9\textwidth]{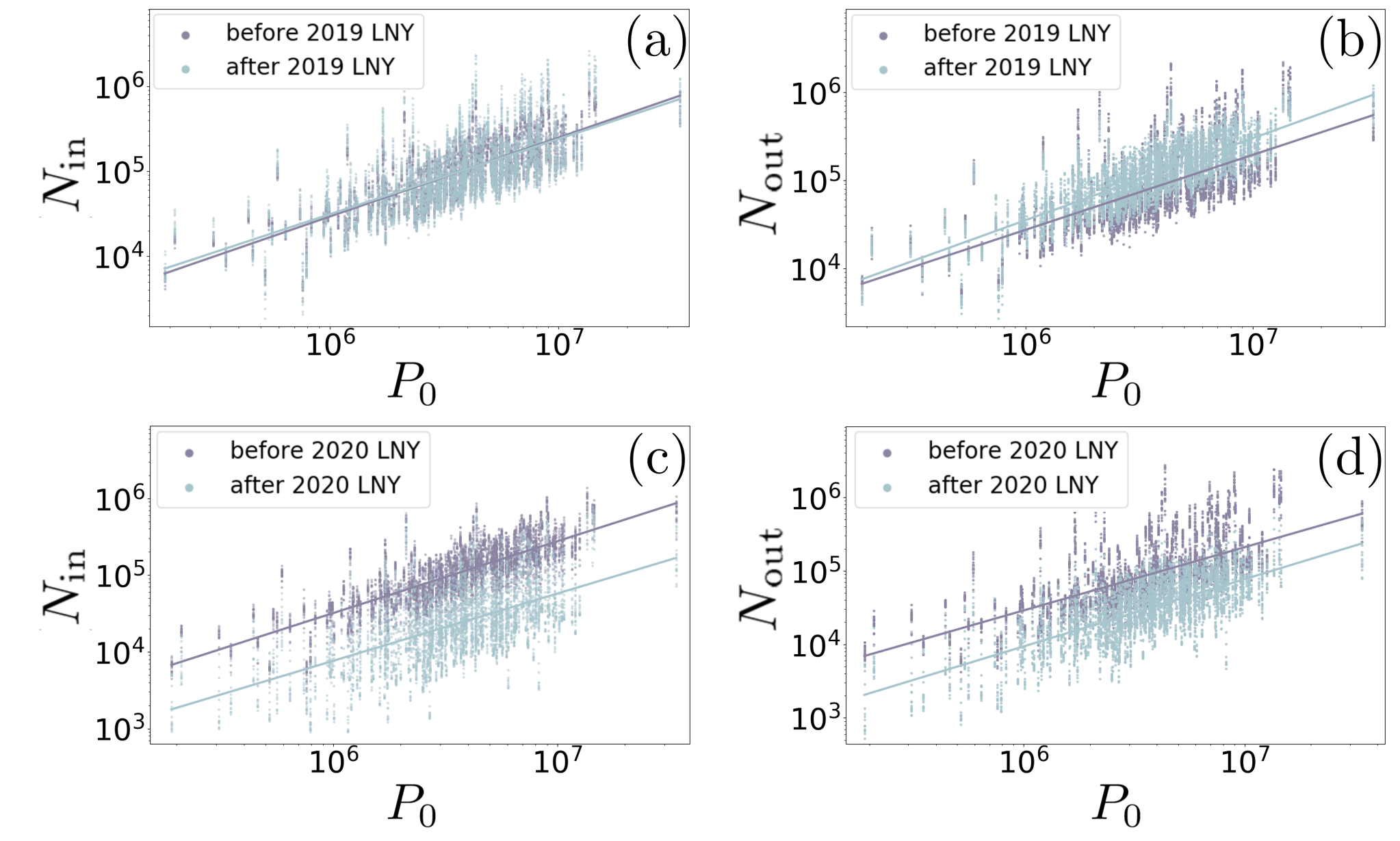}
   \caption{\small Incoming flows (a) and outgoing flows (b) for all cities and days versus city population before and after the 2019 LNY in loglog. Incoming flows (c) and outgoing flows (d) for all cities and days versus city population before and after the 2020 LNY in loglog.}
   \label{Fig_incoming_outgoing population}
 \end{figure}

 \begin{figure}[ht!]
   \centering
   \includegraphics[width=0.9\textwidth]{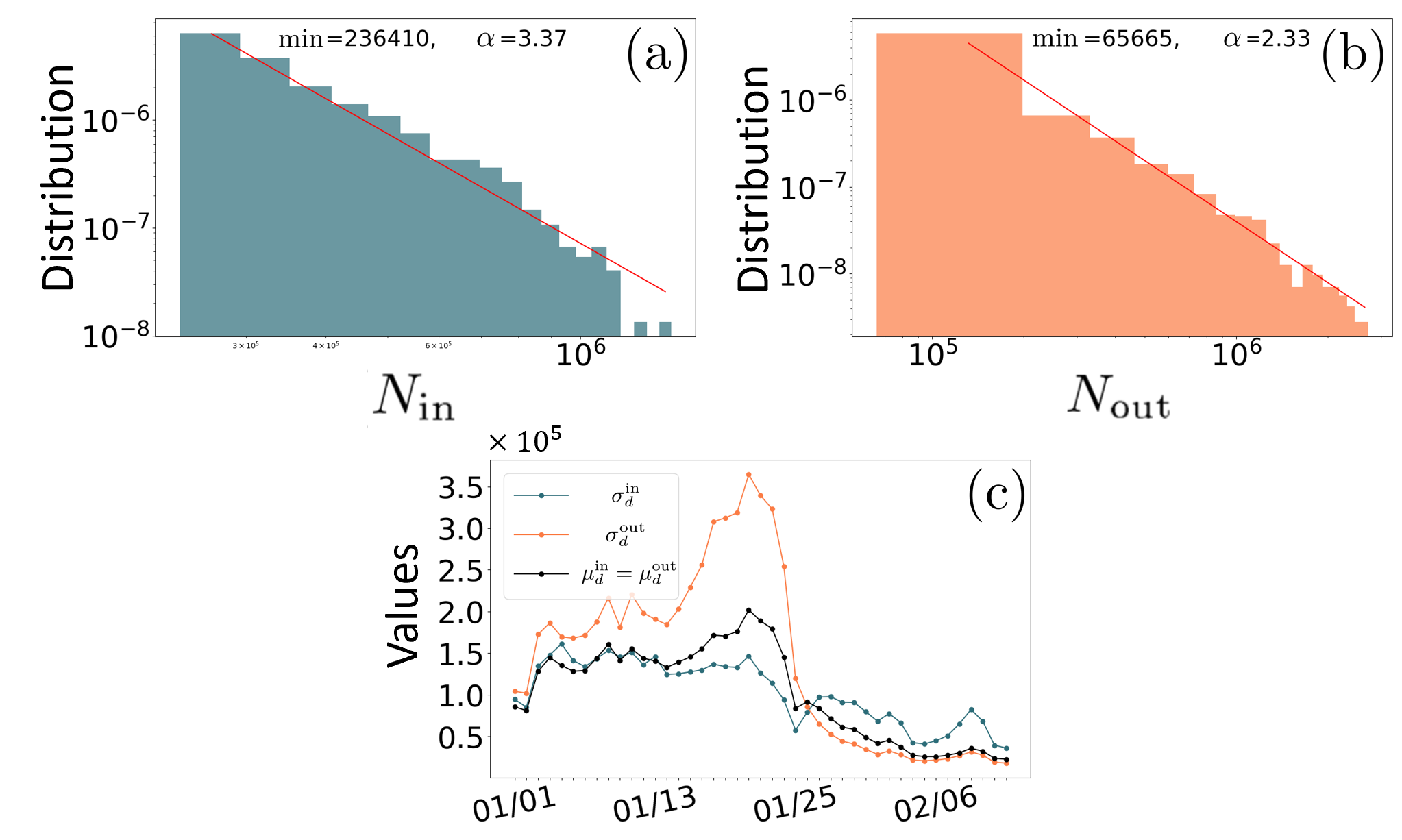}
   \caption{\small Observation of incoming and outgoing flows for 2020. (a) Distribution of all incoming flows in loglog with parameters of power law fitting on the top middle. (b) Distribution of all outgoing flows in loglog with parameters of power law fitting on the top middle. (c) Average and standard deviation of \(N_\text{\text{in}}\) and \(N_\text{\text{out}}\) over cities versus time.}
   \label{Fig_distribution_incoming_outgoing 2020}
 \end{figure}

 \begin{figure}[ht!]
   \centering
   \includegraphics[width=0.9\textwidth]{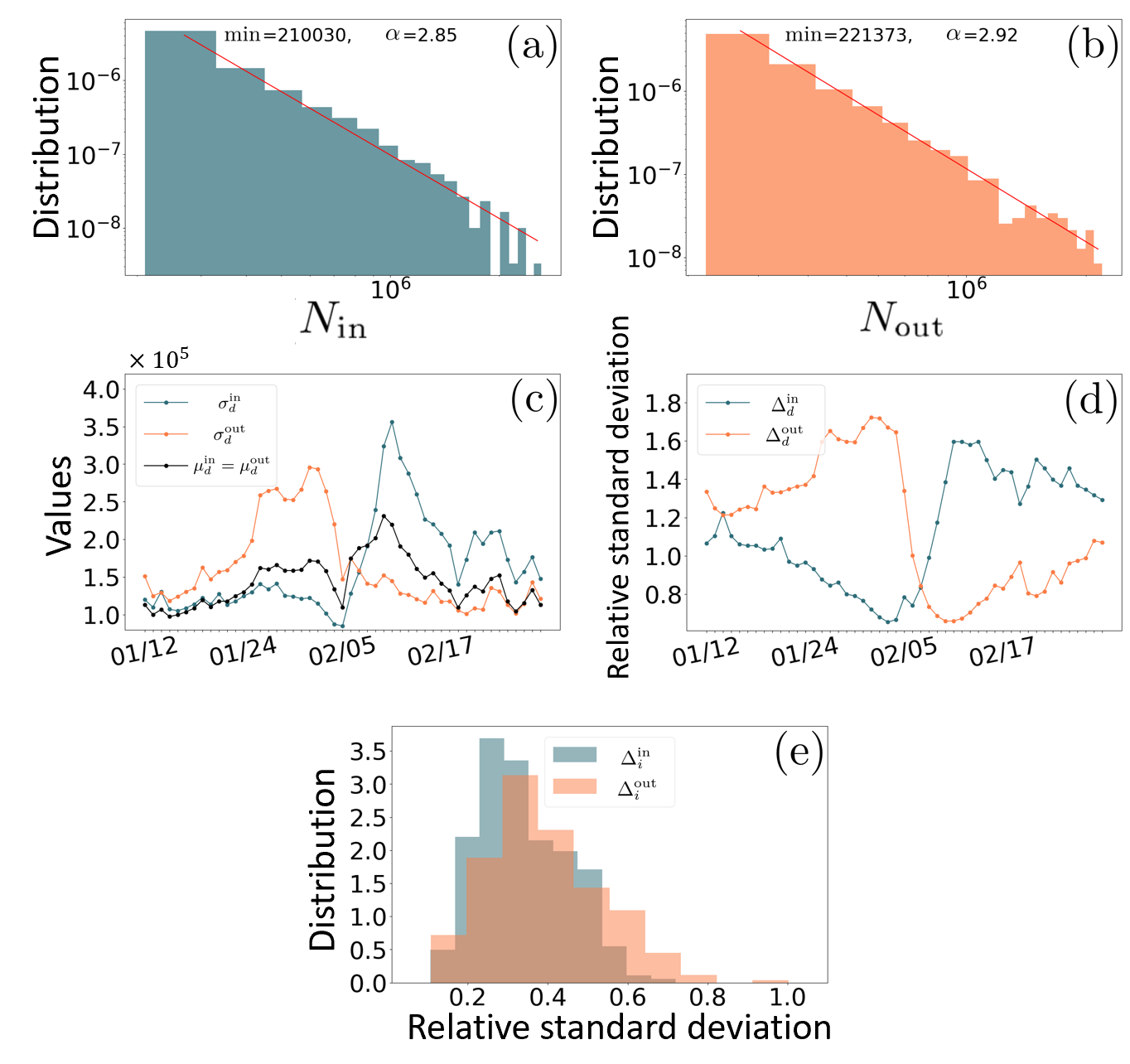}
   \caption{\small Observation of incoming and outgoing flows for 2019. (a) Distribution of all incoming flows in loglog with parameters of power law fitting on the top middle. (b) Distribution of all outgoing flows in loglog with parameters of power law fitting on the top middle. (c) Average and standard deviation of \(N_\text{\text{in}}\) and \(N_\text{\text{out}}\) over cities versus time. (d) Relative standard deviation of \(N_\text{\text{in}}\) and \(N_\text{\text{out}}\) over cities versus time. (e) Distribution of the relative standard deviation of \(N_\text{\text{in}}\) and \(N_\text{\text{out}}\) over time. }
   \label{Fig_distribution_incoming_outgoing 2019}
 \end{figure}

 \begin{figure}[ht!]
   \centering
   \includegraphics[width=0.9\textwidth]{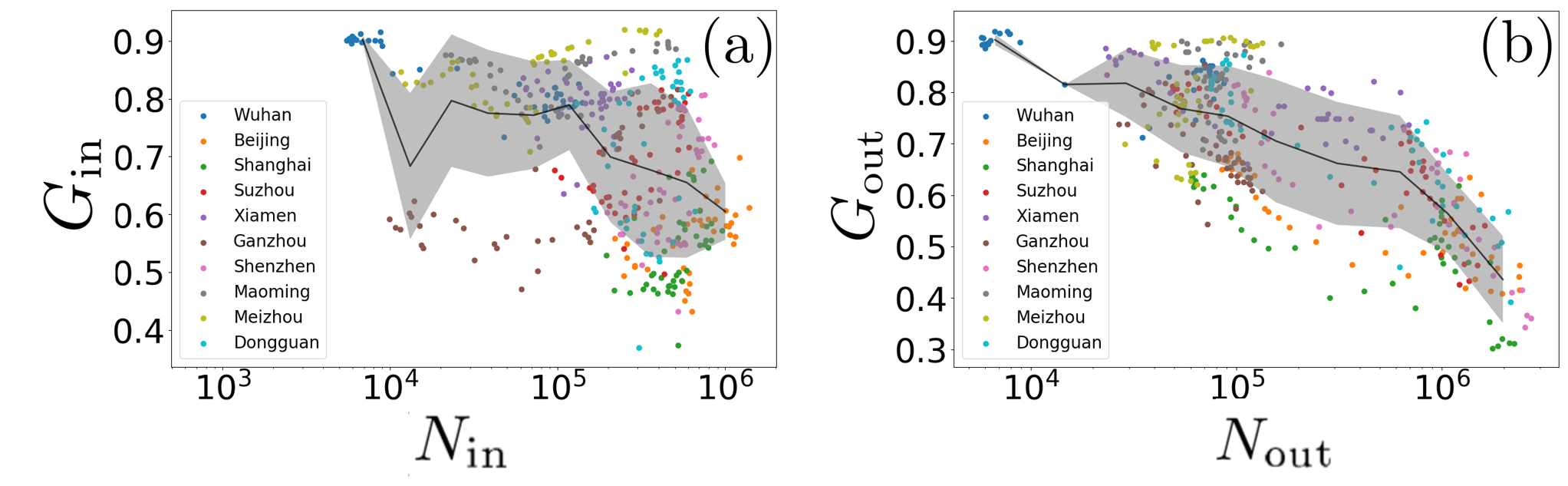}
   \caption{\small (a) \(G_\text{\text{in}}\) versus \(N_\text{\text{in}}\) for some  cities and all days. (b) \(G_\text{\text{out}}\) versus \(N_\text{\text{out}}\) for some  cities and all days.}
   \label{Fig_gini_index_some_cities}
 \end{figure}

 \begin{figure}[ht!]
   \centering
   \includegraphics[width=0.9\textwidth]{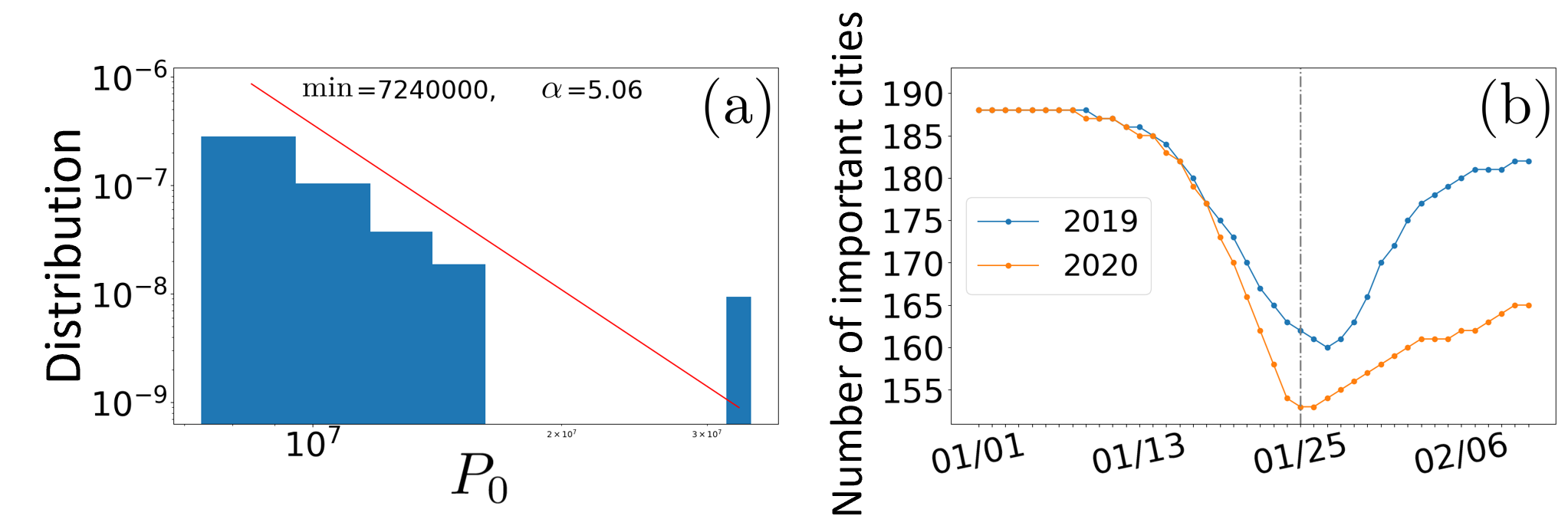}
   \caption{\small (a) Distribution of populations of all cities in loglog. (b) The number of important cities versus time corresponding to 2019 and 2020. We match the time scale for 2019 and 2020 according to LNY (for the sake of clarity, we show on the x-axis the dates for 2020 only). The vertical line highlights the LNY. }
   \label{Fig_population_number_important_cities}
 \end{figure}

 \begin{figure}[ht!]
   \centering
   \includegraphics[width=0.9\textwidth]{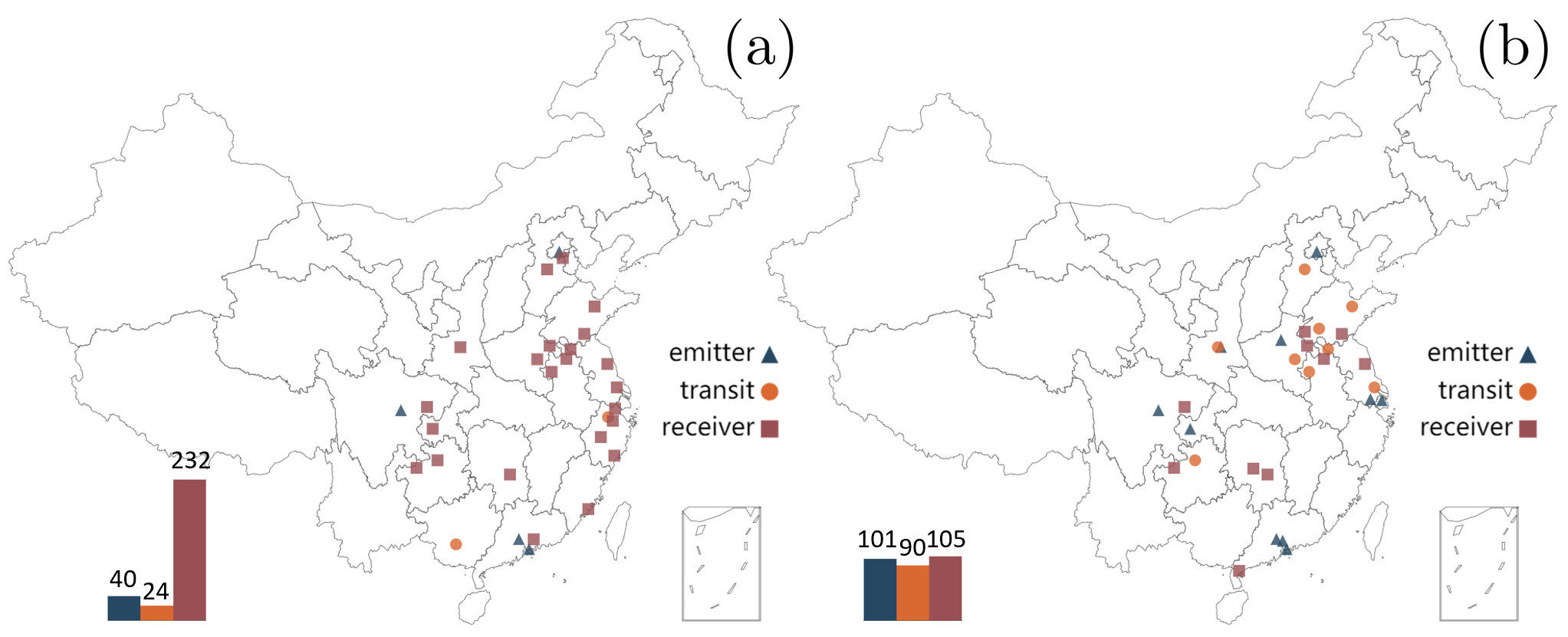}
   \caption{\small (a) Emitter, receiver and transit cities according to the value of \(R(i,1)\) for 2019 with the number of three categories of provinces at the lower left corner. (b) Emitter, receiver and transit cities according to the value of \(R(i,1)\) for 2020 with the number of three categories of provinces at the lower left corner.}
   \label{Fig_average_pendular_ratio_on_map_different_criteria}
 \end{figure}

 \begin{figure}[ht!]
   \centering
   \includegraphics[width=0.9\textwidth]{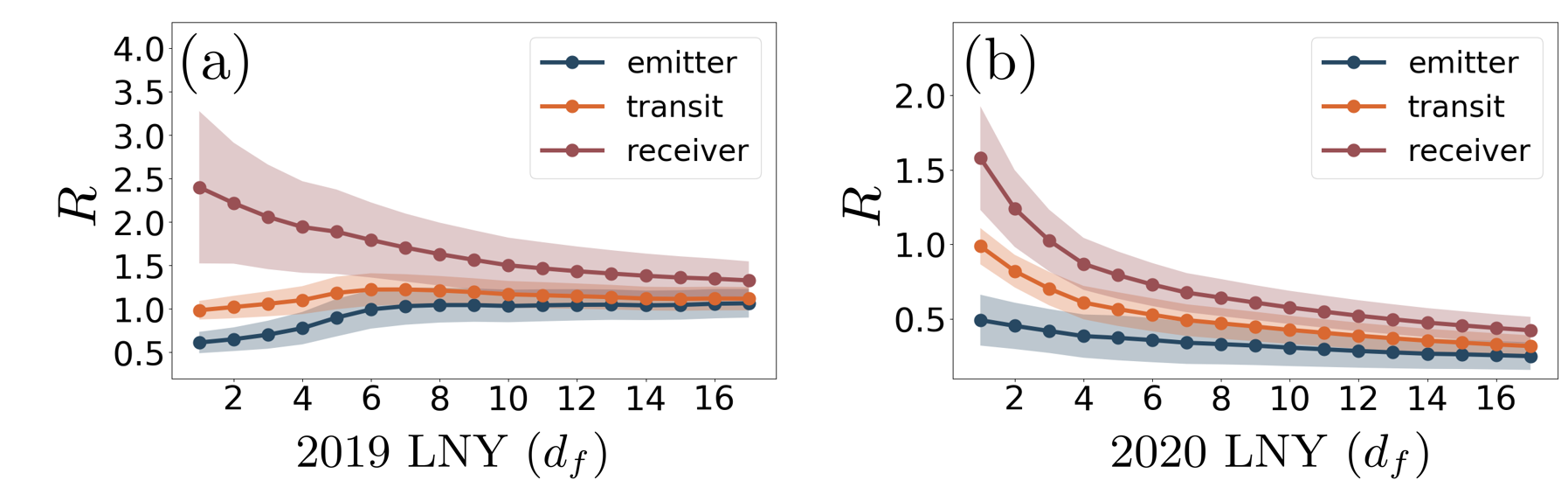}
   \caption{\small (a) Mean value of the pendular ratios over cities according to the classification (receiver, emitter or transit cities) with different criteria versus days from LNY in 2019. The colored areas correspond to one standard deviation. (b) Mean value of the pendular ratios over cities according to the classification with different criteria versus days from LNY in 2020 with shaded areas representing standard deviation.}
   \label{Fig_pendular_different_criteria}
 \end{figure}

 \begin{figure}[ht!]
   \centering
   \includegraphics[width=0.9\textwidth]{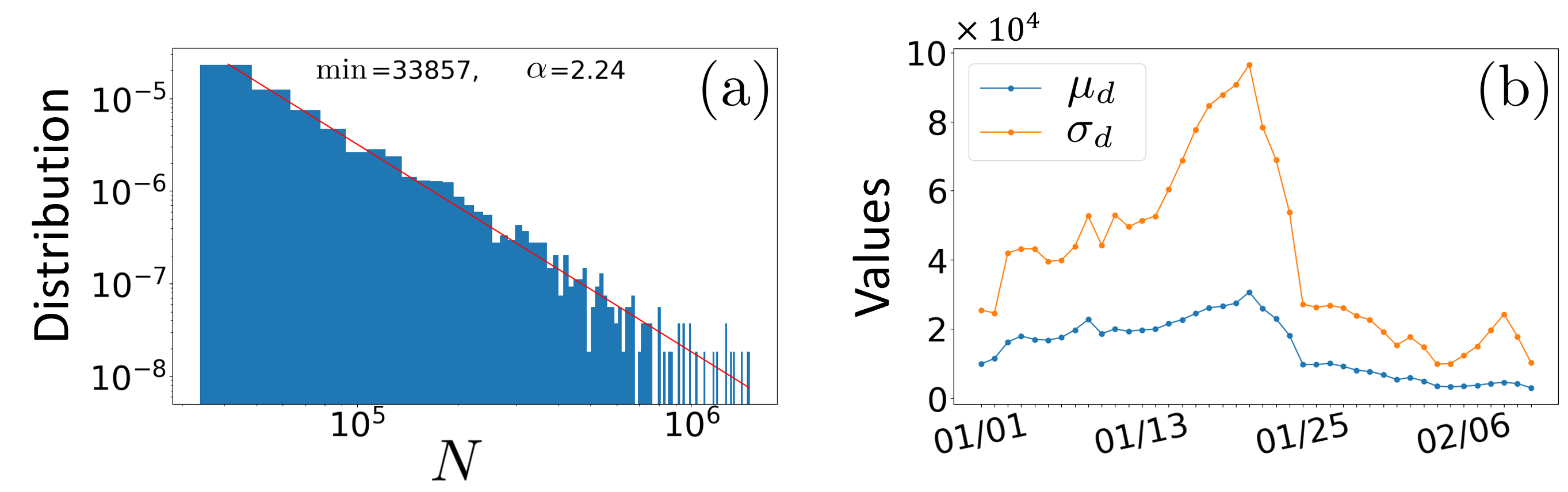}
   \caption{\small (a) Distribution of all traffic flows $N(i,j,d)$ in loglog. The line is a power law fit of the form $N^{-\alpha}$ with exponent $\alpha=2.27$. (b) Average and standard deviation of \(N\) over traffic flows versus time.}
   \label{Fig_distribution_traffic_flow_province}
 \end{figure}

 \begin{figure}[ht!]
   \centering
   \includegraphics[width=0.9\textwidth]{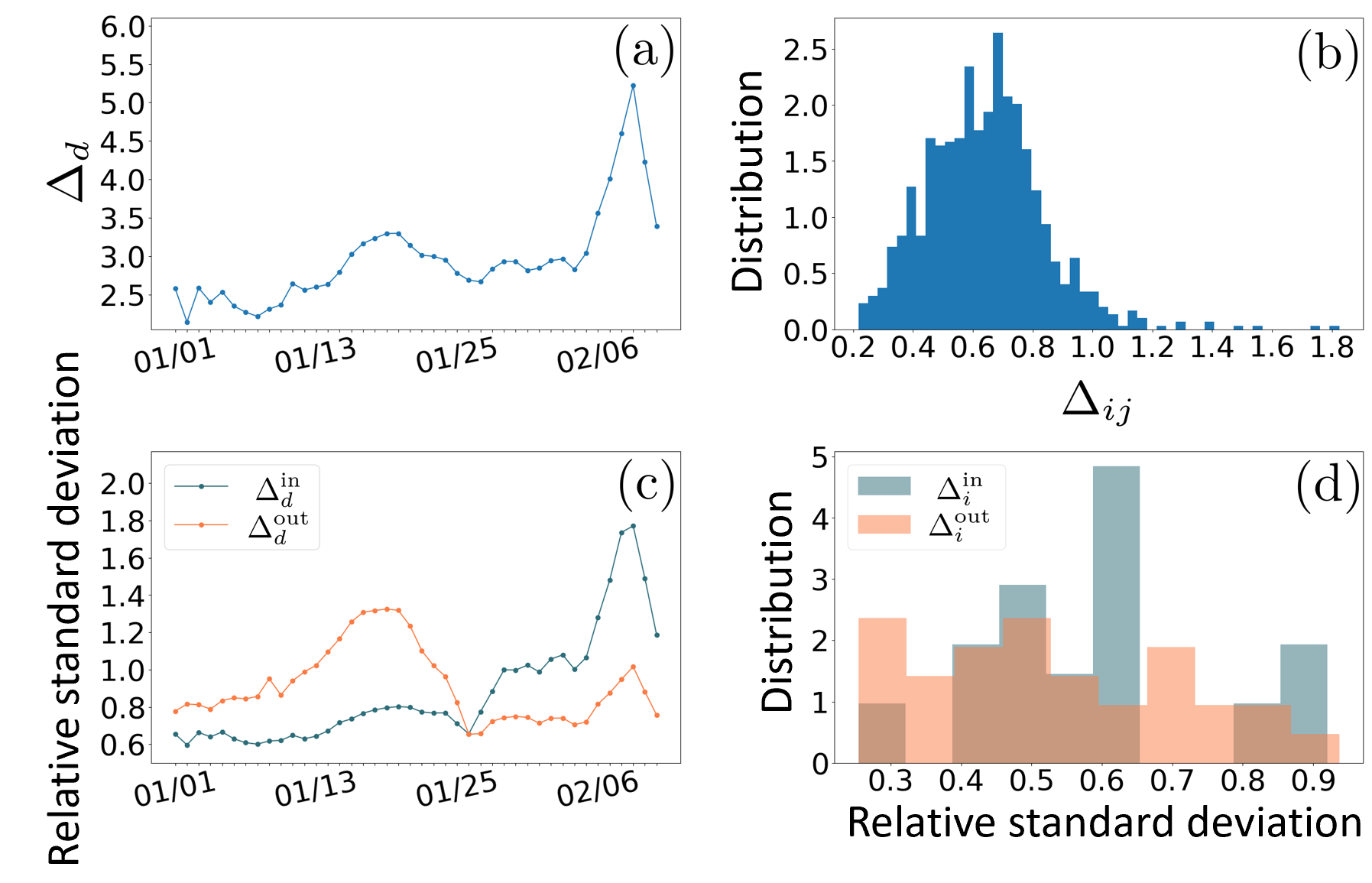}
   \caption{\small  (a) Relative standard deviation of \(N\) over traffic flows versus time. (b) Distribution of the relative standard deviation of \(N\) over time. (c) Relative standard deviation of \(N_\text{\text{in}}\) and \(N_\text{\text{out}}\) over provinces versus time. (d) Distribution of the relative standard deviation of \(N_\text{\text{in}}\) and \(N_\text{\text{out}}\) over time.}
   \label{Fig_temporal_spatial_province}
 \end{figure}

 \begin{figure}[ht!]
   \centering
   \includegraphics[width=0.9\textwidth]{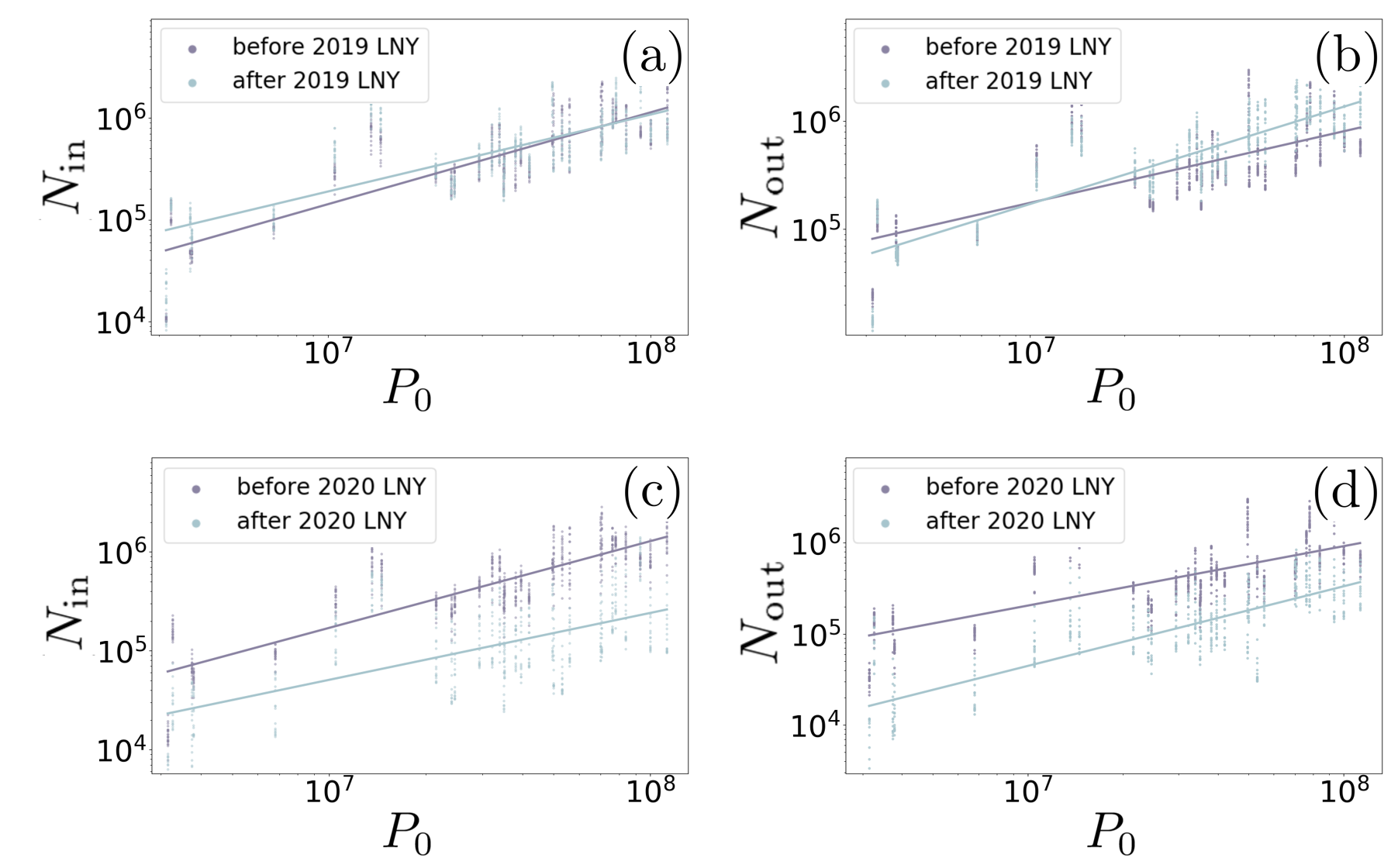}
   \caption{\small Incoming flows (a) and outgoing flows (b) for all provinces and days versus province population before and after the 2019 LNY in loglog. Incoming flows (c) and outgoing flows (d) for all provinces and days versus province population before and after the 2020 LNY in loglog.}
   \label{Fig_incoming_outgoing population_province}
 \end{figure}

 \begin{figure}[ht!]
   \centering
   \includegraphics[width=0.6\textwidth]{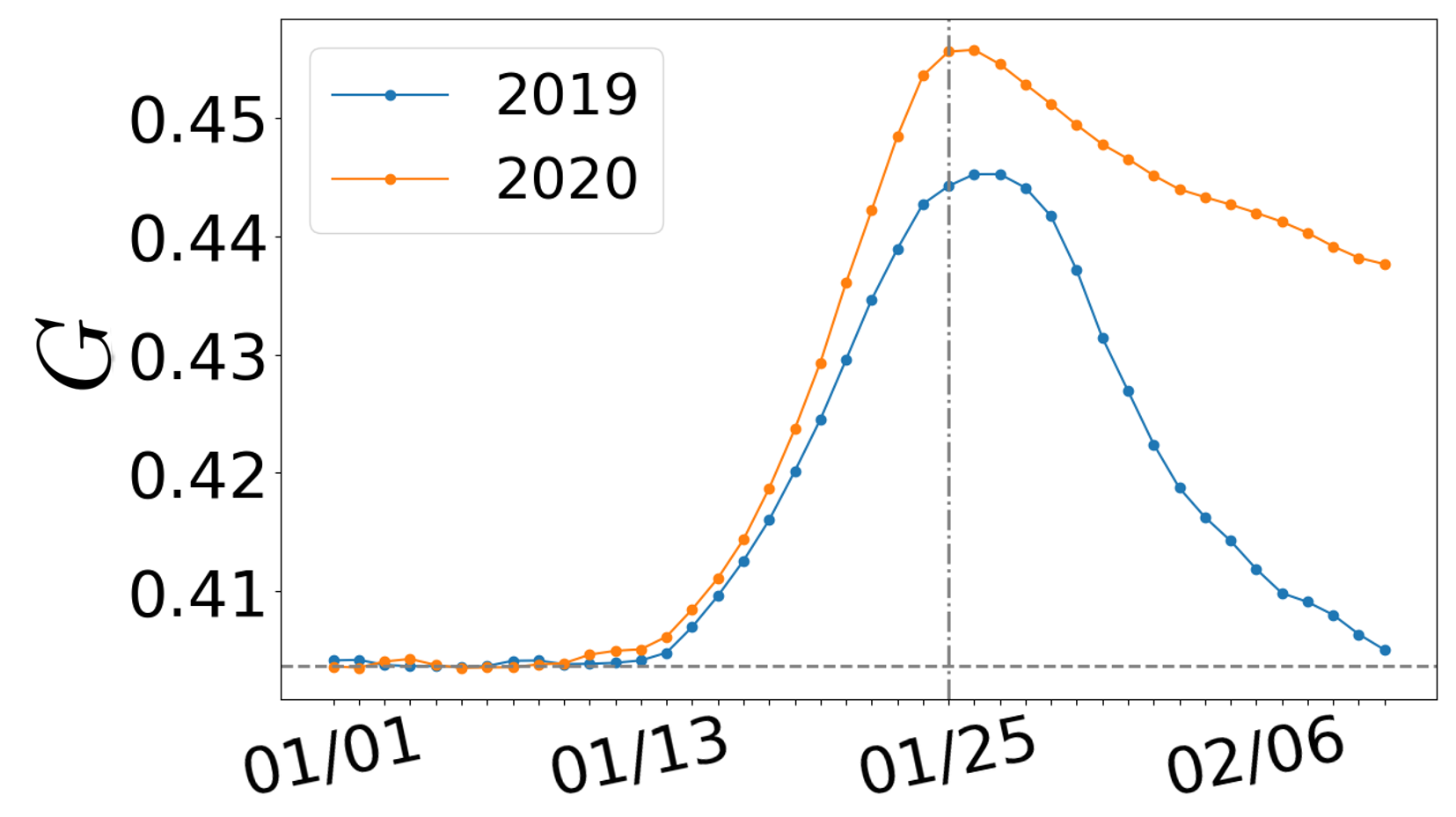}
   \caption{\small Temporal variations during the Spring Festical of the population Gini index for 2019 and 2020. The dotted line represents the value `at rest'.} \label{Fig_gini_t_province}
 \end{figure}

 \begin{figure}[ht!]
   \centering
   \includegraphics[width=0.9\textwidth]{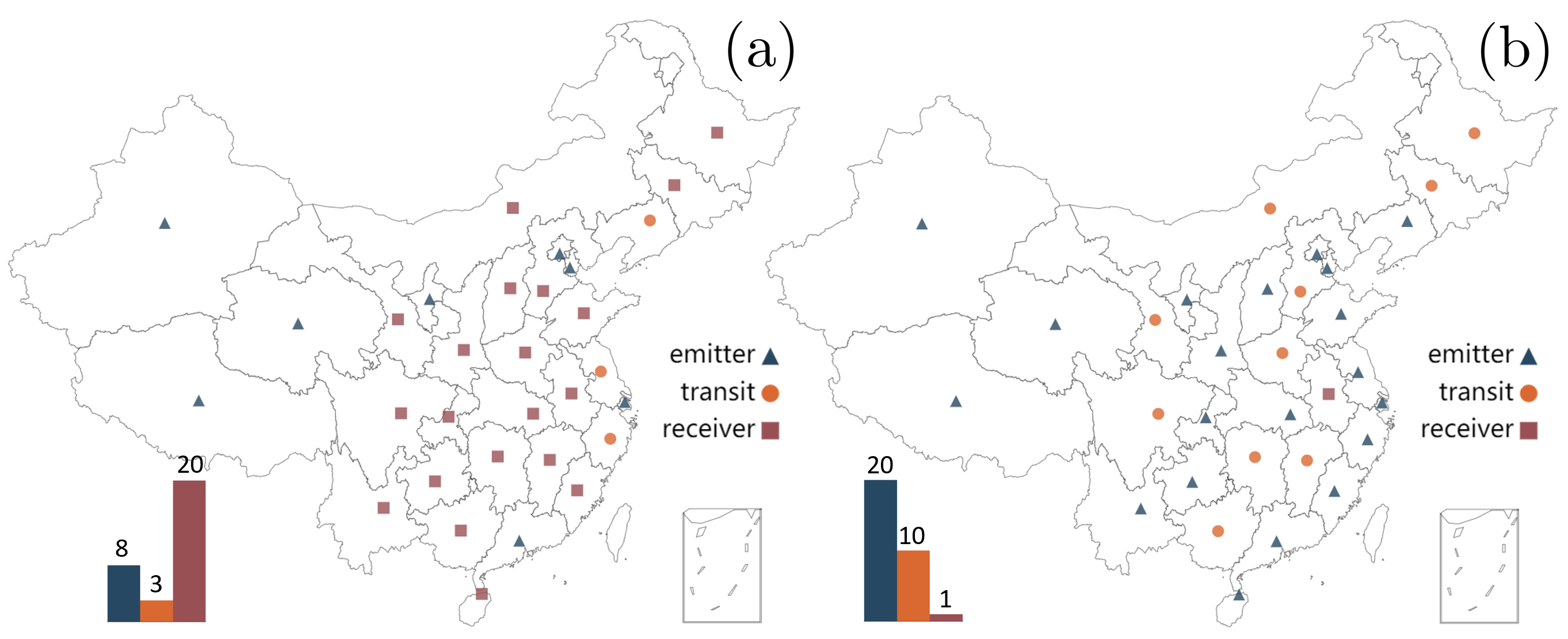}
   \caption{\small (a) Emitter, receiver and transit provinces according to the value of \(R(i,1)\) for 2019 with the number of three categories of provinces at lower left corner. (b) Emitter, receiver and transit provinces according to the value of \(R(i,1)\) for 2020 with the number of three categories of provinces at lower left corner.}
   \label{Fig_average pendular ratio on map_province}
 \end{figure}

 \begin{figure}[ht!]
   \centering
   \includegraphics[width=0.9\textwidth]{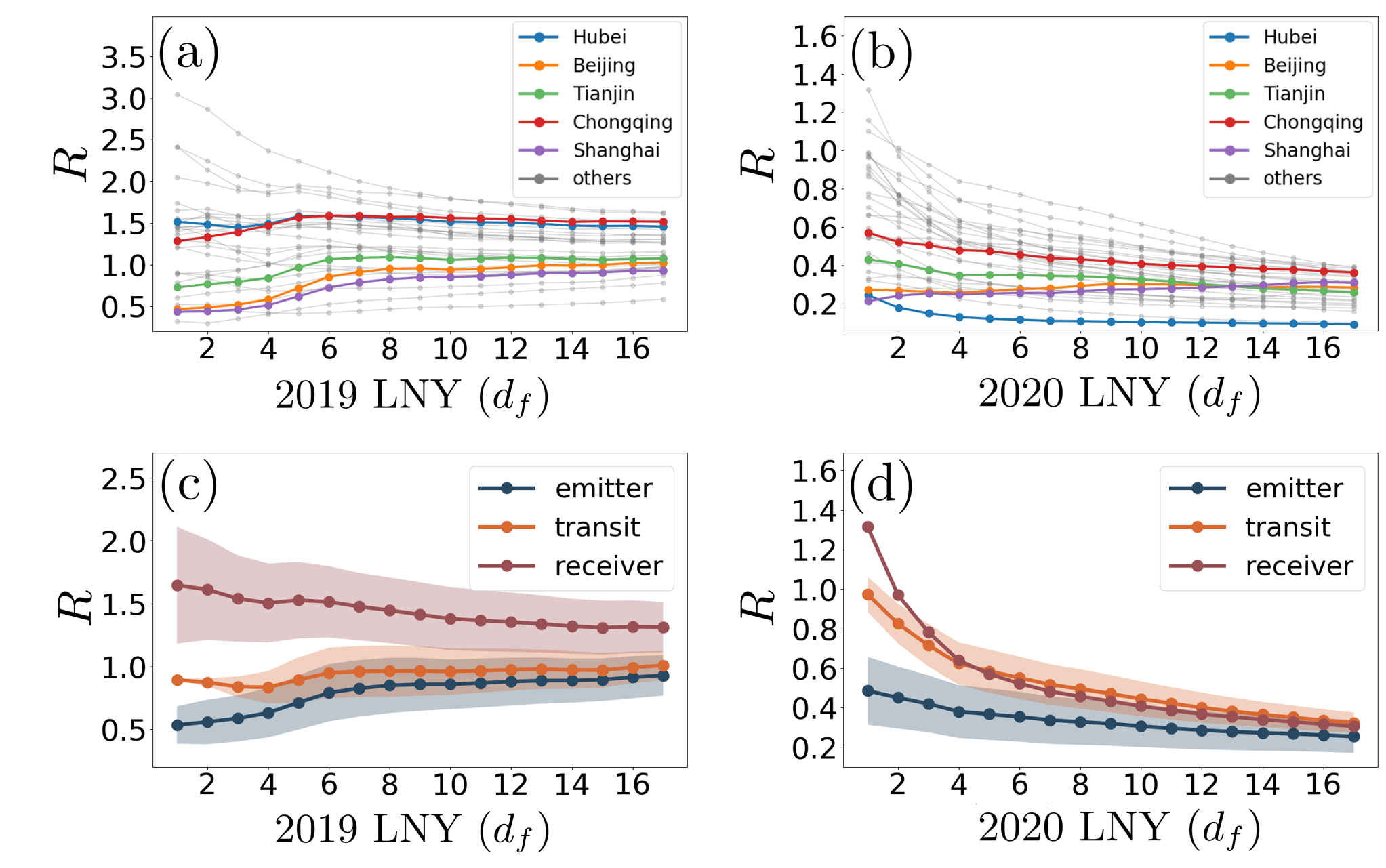}
   \caption{\small Comparison of pendular ratio between 2019 and 2020. (a) Pendular ratio for all provinces versus days $d_f$ from LNY in 2019. We highlighted five important provinces. (b) Pendular ratio for all provinces versus days $d_f$ from LNY in 2020 with highlight of 5 provinces. (c) Mean values of the pendular ratio over provinces according to the classification (receiver, emitter or transit provinces) versus days from LNY in 2019. The colored areas correspond to one standard deviation. (d) Mean value of the pendular ratio over provinces according to the classification versus days from LNY in 2020 with colorbar representing standard deviation.}\label{Fig_pendular_province}
 \end{figure}

\end{document}